\documentclass[11pt]{amsart}
\usepackage{amsmath}
\usepackage{amssymb}
\usepackage{harpoon}
\usepackage[pdftex]{graphicx}
\usepackage{verbatim}
\usepackage{esint}
\usepackage{relsize}
\usepackage[latin1]{inputenc}
\usepackage{xcolor}
\usepackage{tikz-cd}
\theoremstyle{plain}
\newtheorem{theorem}{Theorem}[section]

\theoremstyle{definition}

\renewcommand{\thesignif}

\numberwithin{equation}{section}

\def\inf{\operatorname{inf}}

\theoremstyle{plain}

\numberwithin{equation}{section}

\allowdisplaybreaks

\begin{document}

\title[Cosmic Topology of General Models]{The Topology of General Cosmological Models}

\author[Galloway]{Gregory J. Galloway}
\address{Department of Mathematics\\
University of Miami\\
Coral Gables, FL 33146, USA}
\email{galloway@math.miami.edu}

\author[Khuri]{Marcus A. Khuri}
\address{Department of Mathematics\\
Stony Brook University\\
Stony Brook, NY 11794, USA}
\email{khuri@math.sunysb.edu}

\author[Woolgar]{Eric Woolgar}
\address{Department of Mathematical and Statistical Sciences\\
and Theoretical Physics Institute\\
University of Alberta\\
Edmonton, AB, Canada T6G 2G1}
\email{ewoolgar@ualberta.ca}


\thanks{G. J. Galloway acknowledges the support of NSF Grant DMS-1710808 and Simons Foundation Grant 850541. M. A. Khuri acknowledges the support of NSF Grants DMS-1708798, DMS-2104229, and Simons Foundation Fellowship 681443. E. Woolgar acknowledges the support of a Discovery Grant RGPIN-2017-04896 from the Natural Sciences and Engineering Research Council.}

\begin{abstract}
Is the universe finite or infinite, and what shape does it have? These fundamental questions, of which relatively little is known, are typically studied within the context of the standard model of cosmology where the universe is assumed to be homogeneous and isotropic.
Here we address the above questions in highly general cosmological models, with the only assumption being that the average flow of matter is irrotational. Using techniques from differential geometry, specifically extensions of the Bonnet-Myers theorem, we derive a condition which implies a finite universe and yields
a bound
for its diameter.
Furthermore, under a weaker condition involving the interplay between curvature and diameter, together with the assumption that the universe is finite
(i.e., has closed spatial slices),
we provide a concise list of possible topologies. Namely, the spatial sections then would be either the ring topologies $S^1 \times S^2$, $S^1\tilde{\times}S^2$, $S^1\times\mathbb{RP}^2$, $\mathbb{RP}^3 \#  \mathbb{RP}^3$, or covered by the sphere $S^3$ or torus $T^3$. In particular, under this condition the basic construction of connected sums would be ruled out (save for one), along with the plethora of topologies associated with negative curvature. These results are obtained from consequences of the geometrization of 3-manifolds, by applying a generalization of the almost splitting theorem together with a curvature formula of Ehlers and Ellis.
\end{abstract}

\maketitle

\section{Introduction}
\label{sec1} \setcounter{equation}{0}
\setcounter{section}{1}

The standard model of cosmology, or $\Lambda$CDM model, is based on the  Friedmann-Lema\^{\i}tre-Robertson-Walker (FLRW) solutions of the Einstein equations which describe a universe that is both homogeneous and isotropic, meaning that the set of possible observations does not depend on location or direction. These symmetries characterize the FLRW models, and imply that the 3-dimensional spatial sections (an instant of time) are of constant sectional curvature and that all physical quantities, such as the matter density $\rho$ and Hubble scalar $H$, are constant along the time slices. An interesting feature of these models is the prediction that if
\begin{equation}\label{condition}
8\pi G\rho/c^{2} -3(H/c)^2 +\Lambda>0,
\end{equation}
where $c$, $G$, and $\Lambda$ are respectively the speed of light, and the gravitational and cosmological constant, then the spatial sections are of curvature greater than a positive constant and hence compact (finite). This is the closure result of FLRW cosmologies. It should be noted that in the case when the
inequality of
\eqref{condition} is reversed, the time slices are of zero or negative curvature and can be either open (infinite) or closed  and, if closed, will have nontrivial topology manifested by an infinite fundamental group.

The hypotheses of homogeneity and isotropy are well-motivated when averaged over very large scales, as can be seen by observation \cite{Planck0,Bennett} as well as through the successes of the $\Lambda$CDM model.
Nonetheless, it is of interest to study models that do not require these pillars of the cosmological principle. A well-known theoretical framework that removes the isotropic assumption is that of the Bianchi universe. In \cite{Saadeh1}, a thorough analysis of these models and their relation to the CMB data has been given, and limits on the anisotropic expansion have been found. Still, these models are homogeneous and provide a very specific type of anisotropy. In this paper we drop the postulate of symmetry altogether, and study general cosmological models with the sole assumption that
the average flow of matter is irrotational. This hypothesis guarantees that through any spacetime point, there passes a 3-dimensional submanifold orthogonal to the average flow of matter, which represents the spatial universe at a moment of time.
In this broad setting, we will generalize the closure result described above, giving a condition which implies that these spatial sections are compact. In analogy with \eqref{condition}, the condition is then shown to be satisfied within the margin of error by the observed values of relevant physical quantities. Furthermore, our approach naturally produces an approximate upper bound for the diameter of the spatial slice.

The methods that we use to study the closure property are based on a classical result of differential geometry, known as the Bonnet-Myers theorem (or often simply the Myers theorem) \cite[Theorem 25]{Petersen}. This states that a complete Riemannian manifold with Ricci curvature bounded below by a positive constant must have finite diameter, and also gives
a bound
for this length. In the setting of general cosmological models, a formula of Ehlers and Ellis \cite{Ehlers,Ellis0} implies that a version of the Ricci tensor, referred to as the Bakry-\'{E}mery Ricci curvature, appears on the spatial sections and is coupled to the matter content as well as other aspects of the geometry. An extended version of the Bonnet-Myers theorem \cite{Limoncu} may then be employed to ascertain a diameter bound for time slices. Initial results in this direction were obtained by the first named author in \cite{Galloway-1,Galloway0}, where closure results were found under more stringent conditions that are not necessarily consistent with current observations. Moreover, the suggestion to directly apply the Bonnet-Myers theorem was discussed in \cite{Ellis-1}.

Planck temperature and polarization measurements combined with lensing data show that,
if our universe is FLRW at least,
the closure criterion
is satisfied at approximately an $85\%$ confidence level.
Since the data analysis hypothesizes a background FLRW cosmology, it is possible that the confidence level for the closure criterion may have some dependence on that choice.
But even if the lower bound for the Bakry-\'{E}mery Ricci curvature of spatial sections is not positive
(i.e., so that the universe is possibly closed but the energy density is not sufficient to force the universe to be closed),
it is
reasonable to assume that it is
almost\footnote
{While the quantifier `almost', as used here, may be interpreted in the colloquial sense, other instances have a precise mathematical meaning. Examples of the latter case include `almost abelian group' and `almost splitting' (see \cite{CC}).}
positive. This suggests the application of another result from differential geometry, namely the almost splitting theorem \cite{CC}, which is a refinement of the classical splitting theorem \cite{CheegerGromoll}. The latter result asserts that in the presence of a line (a geodesic which is infinite in both directions and minimizes length between any two of its points), a complete manifold of nonnegative Ricci curvature splits off a Euclidean factor both geometrically and topologically. The almost splitting refinement deals with almost nonnegative Ricci curvature, and while the conclusion is weaker than the original result, strong topological restrictions still follow when applied to compact manifolds. As mentioned above, in the setting of general cosmological models, the Ricci curvature of spatial sections should be replaced by the Bakry-\'{E}mery Ricci curvature. It turns out that the almost splitting result may be extended to the Bakry-\'{E}mery context \cite{GallowayKhuriWoolgar}, and as we will see this is sufficient for cosmological applications. If the spatial sections are indeed compact, the topology is then highly restricted in the sense that the fundamental group must be almost abelian, or rather, it admits an abelian subgroup of finite index.

A major development in topology and geometry came over 15 years ago, when Thurston's geometrization conjecture was confirmed \cite{CaoZhu, KleinerLott, MorganTian, Perelman}. This gives a classification of compact 3-manifolds in terms of 8 model geometries. Consequences of geometrization, when combined with restrictions placed on the fundamental group from the almost splitting theorem, yield a short list of possible topologies for the spatial sections of the universe. More precisely, the time slices can either admit as a covering space the 3-sphere $S^3$ or 3-torus $T^3$, or are covered by $S^1 \times S^2$. In this last case there are only four possibilities. The sections can be the trivial or non-orientable 2-sphere bundle over a circle $S^1\times S^2$ or $S^1\tilde{\times}S^2$, the product of a circle and the 2-dimensional real projective space $S^1\times \mathbb{RP}^2$, or the connected sum of two copies of 3-dimensional real projective space $\mathbb{RP}^3 \# \mathbb{RP}^3$. Interestingly, the last four on this list, which qualitatively may be interpreted in terms of complexity as lying between the first two possibilities, are not considered within the FLRW models because they do not admit metrics of constant curvature. The first two cases are quotients of the sphere or torus, and do arise in the standard model of cosmology. Such quotients of the sphere are known as spherical 3-manifolds, and they are divided into 5 classes related to the Platonic solids, with their classification given in \cite{TS}. The corresponding quotients of the torus are known as Bieberbach manifolds, and there are 10 in total \cite{Wolf}, 6 being orientable and 4 non-orientable. The results of this paper may then be interpreted as reducing the topology question for general cosmological models, which forgo any symmetry hypotheses, to the same question as presented by the FLRW models for positive and zero curvature, plus the four additional candidates arising from the product of a circle with the 2-sphere. Notably, this would eliminate the case of negative curvature with its infinite variety of topologies, as well as the possibility of connected sums, except for one. The connected sum construction is a fundamental method for building new manifolds out of others of the same dimension, and allows for the 3-dimensional classification problem to be focused on the study of basic building blocks through the prime decomposition. From a cosmological perspective, a universe that is a nontrivial connected sum may be interpreted as possessing one or more wormholes between regions having prime 3-manifold topology. Thus, even without the assumption of homogeneity and isotropy, the presence of wormholes is strongly disfavored. While these conclusions, drawn from the almost splitting theorem, apply to closed universes that may not quite achieve closure density, they apply to observations on scales larger than a certain quantity set by the amount that closure density
exceeds the actual mass-energy density. The mathematics asserts that there is such a scale but does not provide a clear method to compute it.

Investigations using the `circles in the sky' method have claimed that many of the Bieberbach manifolds are unlikely candidates for the topology of the universe \cite{Cornishetal}, and robust constraints have been placed on the possible spherical topologies that can be detected \cite{Weeksetal}. Another primary tool that has been used to investigate cosmic topology is the `method of images' \cite{Bond1, Bond2}, which has achieved restrictions in the compact hyperbolic setting. See \cite{Lum} for a survey of results. Furthermore, using the $\Lambda$CDM and Bianchi models, the Planck mission \cite{Planck} has found no significant evidence for nontrivial topology. It should be pointed out, however, that the candidate ring topologies (those covered by $S^1\times S^2$) were not considered in this analysis since they do not arise within the models studied.
Early WMAP data was analysed in \cite{ALS1} and \cite{ALS2}, which ruled out most (but not all) spherical spaces, namely all except for the 3-sphere and two specific quotients of it.
Therefore if the universe is indeed closed, as is suggested by Planck \cite{EllisLarena}, then modulo the ring topologies which have not been previously investigated, the current evidence seems to point towards the 3-sphere as the most likely cosmic topology.
See, however, the very recent analysis of \cite{ABFS} showing evidence for toroidal topology.

We now summarize our results. In Section 2 we give an explicit derivation of the Ehlers-Ellis formula \cite{Ellis}. Using this formula, and without the usual assumptions of homogeneity and isotropy, when critical closure density is 
surpassed
we show that the full list of possible topologies for the universe is 
no larger
than 
%
the spherical spaces
allowed in FLRW models. Furthermore, if the critical density is not quite achieved, we show in Section 3 that 
%
topological restrictions still apply to closed universes that have small diameter. Smallness is measured in terms of the density, though the precise relationship is not known.

\section{The Closure Criterion}
\label{sec2} \setcounter{equation}{0}
\setcounter{section}{2}

\subsection{Background}
We begin by setting the stage for the cosmological models studied here. This will lead to the Ricci curvature formula of Ehlers and Ellis, for spatial slices, from which we will apply an extended version of the Bonnet-Myers theorem to obtain conditions for closure.

Consider a 4-dimensional spacetime $(M^4,\mathbf{g})$ satisfying the Einstein equations
\begin{equation}
\mathbf{R}_{ab}-\frac{1}{2}\mathbf{R}\mathbf{g}_{ab}+\Lambda \mathbf{g}_{ab}=\kappa T_{ab},
\end{equation}
with stress-energy tensor $T$ and $\kappa=8\pi G/c^4$. Assume that there is a smooth unit time-like vector field $u$ which represents the velocity field of the average flow of matter in the universe. This implies the mild restriction that the spacetime is time-orientable, in which $u$ points into the future. Observe that the covariant derivative may be decomposed orthogonally into irreducible parts based on symmetry by
\begin{equation}
\pmb{\nabla}_a u_b=-u_a \dot{u}_b+\frac{1}{3}\theta g_{ab}+\sigma_{ab}+\omega_{ab},
\end{equation}
where $\dot{u}=\pmb{\nabla}_u u$, and $g_{ab}=\mathbf{g}_{ab}+u_a u_b$ is the projection onto the orthogonal complement of $u$. Additionally, if $i,j$ represent directions perpendicular to $u$ then $\theta=\pmb{\nabla}_i u^i$ is the \textit{volume expansion scalar} or $H=\theta/3$ is the \textit{Hubble scalar}, $\sigma_{ij}=\pmb{\nabla}_{\langle i}u_{j\rangle}$ is the \textit{shear tensor}, and $\omega_{ij}=\pmb{\nabla}_{[i}u_{j]}$ is the \textit{vorticity tensor}
with $\sigma_{ab}u^b =\omega_{ab} u^b =0$. Here the notation $\langle$ $\rangle$ indicates tracefree symmetrization, whereas $[$ $]$ indicates anti-symmetrization.

It will also be assumed that the average flow of matter is \textit{irrotational}, that is $\omega=0$. This condition is equivalent to $u\wedge du=0$, and by the Frobenius theorem it guarantees that the distribution of orthogonal subspaces to $u$ is integrable. Thus $u$ is hypersurface orthogonal, so that through any given spacetime point there is a maximal connected 3-dimensional submanifold $M^3$ whose tangent spaces are orthogonal to $u$, and has positive definite induced metric $g$. Physically, $M^3$ may be interpreted as a spatial section, or some instant of time.
Local coordinates $(t,x^1,x^2,x^3)$ may then be introduced in a spacetime neighborhood about any point of $M^3$, such that the metric is expressed as
\begin{equation}\label{1}
\mathbf{g}=-\varphi^2 dt^2+g_{ij}(t,x)dx^i dx^j,
\end{equation}
where $u=\varphi^{-1}\partial_t$. Here $t$ is
a synchronous time coordinate with $t=0$ corresponding to $M^3$, and the $x^i$ are local coordinates on the time slice. Furthermore, letting $\nabla$ denote the `spatial' gradient/covariant derivative, it can be shown that
\begin{equation}\label{011}
X:=\pmb{\nabla}_u u=\nabla\log\varphi=g^{ab}\partial_a \log\varphi \partial_b,
\end{equation}
in particular $X$ is tangent to $M^3$. Note that although the vector field $X$ exists globally, it may not globally be the gradient of a function, as
the expression \eqref{011} is local. In what follows we will use geometrized units where $c=G=1$, so that coordinates carry distance units, time is converted to distance using the speed of light, and $u$ is dimensionless so that $\mathbf{g}(u,u)=-1$.

The setup described above is useful, because it yields natural spatial sections that can then be examined for closure, without imposing symmetry hypotheses. In particular, there is no requirement that the cosmology be close in any sense to an FLRW model.  In order to proceed, we seek to understand positivity properties of the Ricci curvature $R_{ij}$ of the slice $M^3$. From the Einstein equations, together with the Gauss and Codazzi relations, the following formula appears in \cite{Ellis}
(see also \cite{Ehlers} and \cite{Ellis0})
\begin{equation}\label{rformula}
R_{ij}=\nabla_{\langle i}X_{j\rangle} +X_{\langle i} X_{j\rangle}
+\frac{1}{3}\left(2\mu -\frac{2}{3}\theta^2 +|\sigma|^2 +2\Lambda\right)g_{ij} -\dot{\sigma}_{\langle ij\rangle}-\theta \sigma_{ij}+\Pi_{ij},
\end{equation}
where $\dot{\sigma}=\pmb{\nabla}_u\sigma$. The derivation of this formula is not easily discerned from the literature, and for this reason we present one at the end of this section. The remaining quantities appearing in this equation either have been described above or arise from the decomposition of the stress-energy tensor with respect to the flow of matter
\begin{equation}\label{stresse}
\kappa T_{ab}=\mu u_a u_b +q_a u_b + q_b u_a +pg_{ab }+\Pi_{ab},
\end{equation}
where $q_a u^a=0$, $\Pi_{ab}u^b=0$, and $\Pi_{\langle ab\rangle}=\Pi_{ab}$. Here $\mu$ is the energy density, which when expressed in physical units is given by $8\pi G\rho/c^2$. Likewise $q$ represents \textit{momentum density}, $p$ represents the \textit{isotropic pressure}, and $\Pi$ represents the trace-free \textit{anisotropic pressure} all multiplied by $\kappa$; these have dimensions of inverse length squared in geometrized units.
If $q_a =0$ and $\Pi_{ab}=0$ then the matter is a perfect fluid. If in addition $p=0$ then it is a dust, and in some scenarios this pressureless perfect fluid is used to model cold dark matter. However, here, we make no assumption about the type of matter present.

Notice that upon taking a trace of \eqref{rformula} we obtain a simple expression for the scalar curvature of the spatial section
\begin{equation}
R=2\mu-\frac{2}{3}\theta^2 +|\sigma|^2 +2\Lambda.
\end{equation}
Observe that the condition \eqref{condition}, from the closure result of FLRW cosmologies, implies that $R>0$. Nevertheless, this is not sufficient in the current context to conclude that $M^3$ is compact, because this spatial slice is not necessarily of constant curvature. On the other hand, positivity of Ricci curvature, or Bakry-\'{E}mery Ricci curvature is sufficient to obtain the closure property.

\subsection{The Bakry-\'{E}mery Ricci Condition}

Consider an $n$-dimensional Riemannian manifold $(\mathcal{M}^n,g)$, and let $m>0$. Recall that the generalized $m$-Bakry-\'{E}mery Ricci tensor is given by
\begin{equation}\label{bee}
\mathrm{Ric}_{V}^{m}(g)=\mathrm{Ric}(g)
+\frac{1}{2}\mathcal{L}_{V}g-\frac{1}{m}V\otimes V,
\end{equation}
in which (in a slight abuse of notation) $\mathcal{L}_{V}$ denotes Lie differentiation along the vector field metric-dual to the 1-form $V$.
Bakry-\'{E}mery Ricci curvature arises naturally in many contexts, of which one notable instance is the warped product structure of Kaluza-Klein compactification; in the present situation it appears in the Ehlers-Ellis formula \eqref{rformula}, as we make note of below.
A version of the Bonnet-Myers theorem for this type of Ricci tensor was established in \cite{Limoncu}. It states that if the manifold is complete and \begin{equation}
\mathrm{Ric}_{V}^{m}(g)\geq (n+m-1)\lambda g,
\end{equation}
for some constant $\lambda>0$, then the manifold has bounded diameter with the upper bound
\begin{equation}
\mathrm{diam}\left(\mathcal{M}^n \right)\leq \frac{\pi}{\sqrt{\lambda}}.
\end{equation}
Since the manifold is complete with finite diameter we conclude that it is compact, and has finite fundamental group $|\pi_1\left(\mathcal{M}^n \right)|<\infty$. Moreover, in the case that $n=3$, the elliptization portion of the geometrization of 3-manifolds implies that $\mathcal{M}^3$ must be a spherical manifold, and thus is a quotient of $S^3$ by a subgroup of isometries acting properly discontinuously.

These observations may now be applied to general cosmological models. Namely, we may rewrite the Ricci formula for spatial slices \eqref{rformula} to fit within the context of the Bakry-\'{E}mery Ricci curvature, by setting $V=-X$ and $m=1$ to find
\begin{equation}\label{0987}
\mathrm{Ric}_{-X}^1 (g)=\frac{2}{3}\left(\mu+\Lambda +\frac{1}{2}|\sigma|^2 -\frac{1}{3}\theta^2 -\frac{1}{2}\operatorname{div}X-\frac{1}{2}|X|^2\right)g
-\dot{\sigma}-\theta\sigma+\Pi.
\end{equation}
In order to estimate this from below, and make contact with cosmological parameters, we define the \textit{Hubble scalar} $H$, \textit{total matter density} $\Omega_0$, and \textit{dark energy density} $\Omega_{\Lambda}$ as follows
\begin{equation}
H=\frac{1}{3}\theta,\quad\quad\quad
\Omega_0 =\inf_{M^3}\frac{\mu}{3H^2},\quad\quad\quad
\Omega_{\Lambda}=\inf_{M^3}\frac{\Lambda}{3H^2}.
\end{equation}
The remaining terms on the right-hand side of \eqref{0987} should be negligible, and we denote their smallest eigenvalue by
\begin{equation}\label{epsiloneq}
\varepsilon=\inf_{x\in M^3}\min_{ \substack{w\in T_{x}M^3 \\ |w|=1}}
\left[\left(\frac{|\sigma|^2}{6H^2} -\frac{\operatorname{div}X+|X|^2}{6H^2}\right)g
-\frac{\dot{\sigma}}{2H^2}-\frac{3\sigma}{2H}+\frac{\Pi}{2H^2}\right](w,w).
\end{equation}
It follows that
\begin{equation}\label{yt6}
\mathrm{Ric}_{-X}^1 (g)\geq 2 \left(\Omega_0+\Omega_{\Lambda}-1+\varepsilon\right)H^2 g.
\end{equation}
By applying the generalized Bonnet-Myers theorem we obtain a closure result and diameter
bound
for the universe. The \textit{Hubble constant} will be set as $H_0 =\inf_{M^3} H$.

\begin{theorem}\label{closuret}
Consider a 4-dimensional spacetime satisfying the Einstein equations with matter, and assume that there exists a smooth unit time-like irrotational vector field which represents the velocity field of the average flow of matter in the universe. Let $M^3$ be a complete spatial section orthogonal to the average flow or matter with positive Hubble constant $H_0 >0$. If the sum of the matter and dark energy densities is sufficiently large so that
\begin{equation}\label{c}
\Omega=\Omega_0 +\Omega_{\Lambda}>1-\varepsilon,
\end{equation}
then the following properties hold.

\begin{itemize}
\item [(i)]  The universe is closed, that is, $M^3$ is compact.\smallskip

\item [(ii)] The spatial section $M^3$ is a spherical 3-manifold, namely, it is a quotient of $S^3$ by a subgroup of isometries acting properly discontinuously.\smallskip

\item [(iii)] The diameter of the universe satisfies
\begin{equation}\label{6yt7f}
\mathrm{diam}\left(M^3 \right)\leq
\frac{\pi}{H_0}\sqrt{\frac{3}{2(\Omega -1 +\varepsilon)}}.
\end{equation}
\end{itemize}
\end{theorem}

How large might the right-hand side of \eqref{6yt7f} be? According to
\cite[(47a) page 40]{Planck1}, if one assumes the universe to be FLRW then data from the CMB temperature and polarization anisotropies together with lensing yield
\begin{equation}
\label{eq3.4}
\Omega=1.0106  \pm 0.0065
\end{equation}
at $68\%$ confidence yielding an upper bound for the diameter of the universe
\begin{equation}
\mathrm{diam}\leq 1.2\times 10^{28} \mathrm{m}
\end{equation}
at $85\%$ confidence. This is larger than estimates obtained using traditional models, but depends on many assumptions, including that Planck data analyzed using an assumed FLRW background will not yield an appreciably different result if the universe is not (nearly) FLRW, and that the estimate \eqref{eq3.4} applies to the entire universe and not just the observable portion. Yet there is much room for error, since this estimate is well beyond the distance from Earth to the edge of the observable universe, which in \cite{Halpern} is approximated to be $4.26 \times 10^{26} \mathrm{m}$.

It should be noted, however, that the fundamental question of whether closure density is actually achieved remains undecided and is an area of intense investigation. As the theorem indicates, the possible topologies in this regime are tightly constrained, even far from the setting of homogeneous and isotropic cosmologies. On the other hand, if closure density is not achieved then the almost splitting result could apply, which implies restrictive yet milder topological constraints that are discussed in the next section.

\subsection{The Ehlers and Ellis Spatial Ricci Formula}

Here we present justification for the Ricci formula \eqref{rformula} associated with spatial sections. This formula appears in \cite{Ellis} (see also \cite{Ehlers} and \cite{Ellis0}), although a detailed derivation does not seem to be recorded in the literature.

First observe that the second fundamental form of a time slice is given by $A_{ij}=\langle\nabla_i u,\partial_j\rangle$, and by the Gauss equations
\begin{equation}
\mathbf{R}_{ijkl}=R_{ijkl}+A_{ik}A_{jl}-A_{il}A_{jk},
\end{equation}
where $\mathbf{R}_{ijkl}$ and $R_{ijkl}$ denote the curvature tensors of $\mathbf{g}$ and $g$, respectively, while $i$, $j$, $k$, and $l$ indicate directions tangent to the slice. Taking traces produces
\begin{equation}
\mathbf{R}_{jl}+\mathbf{R}_{ujul}=R_{jl}+\theta A_{jl}-A_{li}A^{i}_j,\quad\quad\quad
\mathbf{R}+2\mathbf{R}_{uu}=R+\theta^2 -|A|^2,
\end{equation}
with $\theta=g^{ij}A_{ij}$. By setting $X=\pmb{\nabla}_u u$, the curvature tensor components are given by
\begin{equation}\label{--}
\mathbf{R}_{ujul}=\nabla_{j}X_{l}-\langle\pmb{\nabla}_u \pmb{\nabla}_j u,\partial_l \rangle
+X_j X_l.
\end{equation}
To see this note that
\begin{equation}
\mathbf{R}_{ujul}=\langle\mathbf{R}(\partial_j,u)u,\partial_l\rangle
=\langle\pmb{\nabla}_j \pmb{\nabla}_u u-\pmb{\nabla}_u \pmb{\nabla}_j u-\pmb{\nabla}_{[\partial_j, u]}u,\partial_l\rangle,
\end{equation}
and combine it with
\begin{equation}
\left[\partial_{j},{u}\right]=\left[\partial_{j},{\phi}^{-{1}}\partial_{t}\right]
=-{\frac{{\partial_{j}\phi}}{{\phi}}}{u}=-X_j u,
\end{equation}
as well as
\begin{equation}
-\langle\nabla_{\left[\partial_{j},{u}\right]}{u},\partial_{l}\rangle
=X_j \langle \pmb{\nabla}_u u , \partial_l \rangle= X_j X_l.
\end{equation}
It follows that
\begin{align}
\begin{split}
R_{jl}=&\mathbf{R}_{jl}+\mathbf{R}_{ujul}-\theta A_{jl}+A_{jl}^2\\
=&\mathbf{R}_{jl}+\nabla_j X_l +X_j X_l -\langle\pmb{\nabla}_u \pmb{\nabla}_j u,\partial_l \rangle -\theta A_{jl} +A_{jl}^2.
\end{split}
\end{align}

Next, by evaluating the Einstein equations along tangential direction to the slice we obtain
\begin{equation}
\mathbf{R}_{jl}=\frac{1}{2}\mathbf{R} g_{jl}+\kappa T_{jl}-\Lambda g_{jl}.
\end{equation}
Below it will be shown that
\begin{equation}\label{ba}
\langle\pmb{\nabla}_u \pmb{\nabla}_j u,\partial_l\rangle=\dot{A}_{jl}+A_{jl}^2,
\end{equation}
and therefore
\begin{equation}\label{100}
R_{jl}=\frac{1}{2}\mathbf{R}g_{jl}+\kappa T_{jl}-\Lambda g_{jl}+\nabla_j X_l
+X_j X_l -\theta A_{jl}-\dot{A}_{jl}.
\end{equation}
To confirm \eqref{ba} observe that
\begin{align}\label{uhb}
\begin{split}
\langle\pmb{\nabla}_u \pmb{\nabla}_j u,\partial_l\rangle=&u\langle\pmb{\nabla}_j u,\partial_l\rangle
-\langle\pmb{\nabla}_j u,\pmb{\nabla}_u \partial_l\rangle\\
=& u\langle\pmb{\nabla}_j u,\partial_l\rangle-A_{jl}^2-\langle\pmb{\nabla}_j u,[u,\partial_l]\rangle\\
=&u\langle\pmb{\nabla}_j u,\partial_l\rangle-A_{jl}^2,
\end{split}
\end{align}
since
\begin{equation}
\langle\pmb{\nabla}_j u,[u,\partial_l]\rangle=-X_l
\langle\nabla_j u,u\rangle=0.
\end{equation}
Furthermore direct computation shows that
\begin{align}
\begin{split}
u\langle\pmb{\nabla}_j u,\partial_l\rangle=&uA_{jl}\\
=&\left(\pmb{\nabla}_u A\right)(\partial_j,\partial_l)+A(\pmb{\nabla}_u \partial_j,\partial_l)
+A(\partial_j,\pmb{\nabla}_u \partial_l)\\
=&\dot{A}_{jl}+A(\pmb{\nabla}_u \partial_j,\partial_l)
+A(\partial_j,\pmb{\nabla}_u \partial_l),
\end{split}
\end{align}
and
\begin{align}
\begin{split}
A(\pmb{\nabla}_u \partial_j,\partial_l)=&\varphi^{-1}A(\pmb{\nabla}_t \partial_j,\partial_l)\\
=&\varphi^{-1}A(\pmb{\nabla}_j \partial_t,\partial_l)\\
=&A(\pmb{\nabla}_j u,\partial_l)-(\partial_j \varphi^{-1})A(\partial_t,\partial_l)\\
=&A_{j}^{k}A_{kl}-(\partial_j \varphi^{-1})A(\partial_t,\partial_l),
\end{split}
\end{align}
and
\begin{equation}
A(\partial_t,\partial_l)=\langle\pmb{\nabla}_t u,\partial_l\rangle
=-\langle u,\pmb{\nabla}_t \partial_l\rangle
=-\langle u,\pmb{\nabla}_l \partial_t \rangle
=\langle \pmb{\nabla}_l u,\partial_t\rangle
=\varphi\langle\pmb{\nabla}_l u,u\rangle=0,
\end{equation}
so that
\begin{equation}\label{plm}
u\langle\pmb{\nabla}_j u,\partial_l\rangle=\dot{A}_{jl}+2A_{jl}^2.
\end{equation}
The desired conclusion \eqref{ba} now follows from \eqref{uhb} and \eqref{plm}.

Consider the trace-free second fundamental form $\sigma_{ij}=A_{ij}-\frac{1}{3}\theta g_{ij}$.
Differentiating produces
\begin{equation}\label{sigmadot}
\dot{\sigma}_{ij}=\dot{A}_{ij}-\frac{1}{3}\dot{\theta}g_{ij},
\end{equation}
since
\begin{equation}
\dot{g}_{ij}=(\pmb{\nabla}_u g)(\partial_i,\partial_j)
=(\pmb{\nabla}_u \mathbf{g})(\partial_i,\partial_j)+\pmb{\nabla}_u(u\otimes u)(\partial_i,\partial_j)=0.
\end{equation}
Furthermore, recall the decomposition of the stress-energy tensor \eqref{stresse} to find
\begin{equation}
\mathrm{Tr}_{\mathbf{g}}\kappa T=-\mu+3p.
\end{equation}
The Raychaudhuri equation then becomes
\begin{equation}
\dot{\theta}=-|\sigma|^2 -\frac{1}{3}\theta^2 -\frac{1}{2}(\mu+3p)+\Lambda+\mathrm{div}X +|X|^2.
\end{equation}
Combining this with \eqref{sigmadot} gives
\begin{align}\label{jkl}
\begin{split}
\dot{A}_{jl}=&\dot{\sigma}_{jl}+\frac{1}{3}\dot{\theta}g_{jl}\\
=&\dot{\sigma}_{jl}+\frac{1}{3}\left(-|\sigma|^2-\frac{1}{3}\theta^2 -\frac{1}{2}(\mu+3p)+\Lambda+\mathrm{div}X+|X|^2\right)g_{jl}.
\end{split}
\end{align}

Now take a trace of the Einstein equations to compute the spacetime scalar curvature
\begin{equation}\label{trace}
\mathbf{R}=4\Lambda-\mathrm{Tr}_{\mathbf{g}}\kappa T=4\Lambda
+\mu-3p.
\end{equation}
Use this, and insert \eqref{jkl} into \eqref{100} to produce
\begin{align}
\begin{split}
R_{jl}=&\nabla_j X_l +X_j X_l +\kappa T_{jl}-\theta A_{jl} -\dot{\sigma}_{jl}\\
&+\frac{1}{3}\left(|\sigma|^2 +\frac{1}{3}\theta^2 +2\mu-3p
-\mathrm{div}X-|X|^2 +2\Lambda\right)g_{jl}.
\end{split}
\end{align}
Since
\begin{equation}
\kappa T_{jl}=\kappa T_{\langle jl\rangle}+\frac{1}{3}\left(\mathrm{Tr}_{g}\kappa T\right)g_{jl}
=\Pi_{jl}+p g_{jl},
\end{equation}
we have
\begin{equation}\label{-=}
R_{jl}=\nabla_{\langle i}X_{l\rangle}+X_{\langle j}X_{l\rangle}+\Pi_{ij}-\theta A_{jl} -\dot{\sigma}_{jl}
+\frac{1}{3}\left(2\mu +\frac{1}{3}\theta^2 +|\sigma|^2 +2\Lambda\right)g_{jl}.
\end{equation}
Next note that
\begin{equation}\label{++}
\theta A_{jl} =\theta \sigma_{jl}+\frac{\theta^2}{3}g_{jl},
\end{equation}
and
\begin{equation}\label{+++}
\dot{\sigma}_{jl}=\dot{\sigma}_{\langle jl\rangle}
+\frac{1}{3}\left(g^{ik}\dot{\sigma}_{ik}\right) g_{jl}=\dot{\sigma}_{\langle jl\rangle},
\end{equation}
as $\dot{\sigma}$ is trace-free. To see this last point, observe that the full spacetime trace of the shear tensor vanishes, so that
\begin{equation}
0 = \partial_t ({\bf g}^{ab}\sigma_{ab}) =
{\bf g}^{ab}\pmb{\nabla}_t\sigma_{ab} =
g^{ij}\pmb{\nabla}_t \sigma_{ij}  = \phi g^{ij} \dot{\sigma}_{ij},
\end{equation}
where we have used that $\nabla_{t}\sigma(\partial_t,\partial_t) = 0$. Combining \eqref{++} and \eqref{+++} with \eqref{-=}, yields the desired Ricci formula \eqref{rformula}.

\section{Cosmic Topology of a Closed Universe}
\label{sec3} \setcounter{equation}{0}
\setcounter{section}{3}

In Section \ref{sec2} we saw that positive Bakry-\'{E}mery Ricci curvature of spatial slices may be used to obtain a closure result and diameter
bound.
%
This led to a highly restrictive list of possible universe topologies consisting of spherical space forms. It is nevertheless possible that this lower bound could be slightly negative. In this situation the Bonnet-Myers theorem and its generalizations are not applicable. However, the almost splitting theorem \cite{CC} is in fact specifically designed for this scenario. As discussed in the introduction, the almost splitting result gives conditions under which the manifold almost splits-off a Euclidean factor. Here we will be interested in the topological conclusions that follow from this almost splitting, 
under the \emph{assumption} that the universe is spatially closed. 
While the original theorem was established for pure Ricci curvature, a recent extension \cite[Theorem 1.3]{GallowayKhuriWoolgar} has been established for Bakry-\'{E}mery Ricci curvature that is applicable to the cosmological context. The next result translates this theorem to the current setting.

\begin{theorem}\label{almostabelian}
Consider a 4-dimensional spacetime satisfying the Einstein equations, and assume that there exists a smooth unit time-like irrotational vector field which represents the velocity field of the average flow of matter in the universe. Let $M^3$ be a compact spatial section orthogonal to the average flow of matter with Hubble scalar $H_0$. There exists a sufficiently small $\delta>0$ dependent on diameter, volume, and size of $X$, such that if
\begin{equation}\label{7890}
\left(\Omega-1+\varepsilon\right)H_0^2 \geq -\delta,
\end{equation}
then the fundamental group $\pi_1 \left(M^3 \right)$ is almost abelian. In particular, the spatial slice admits a finite cover whose fundamental group is abelian.
\end{theorem}

The restrictions on the fundamental group imposed by this theorem, together with consequences of the geometrization theorem for 3-manifolds, yields a list of possible topologies for the universe that are detailed below. An important consequence is that relatively small negative (Bakry-\'{E}mery Ricci) curvature,
which is not inconsistent with the range of possibilities implied by current measurements, does not permit spatial sections to have the topology of a hyperbolic manifold; here relative smallness is interpreted with respect to diameter.
By way of contrast, in
the standard model of cosmology, any amount of negative curvature implies that spatial sections must have the topology of a hyperbolic manifold.

A primary hypothesis of Theorem \ref{almostabelian} is the smallness requirement of the constant $\delta$. The degree to which $\delta$ must be small for the conclusions to be valid relies on the diameter and volume of $M^3$, as well as on $X$. On the other hand, we may obtain a rough estimate of a lower bound for $\delta$ guaranteeing applicability to the observable universe,
in a manner similar to the diameter estimation.
We may assume with
some confidence that $H_0 > 5\times 10^{-27} m^{-1}$. Planck CMB spectra combined with lensing and BAO data \cite[Fig. 26]{Planck1} (see also \cite[Fig. 1]{EG00}) imply, for an FLRW universe at least (this being an implicit assumption in the data analysis), that
\begin{equation}\label{tytyi}
\left(\Omega-1+\varepsilon\right)H_0^2 \geq -1.25 \times 10^{-55} \mathrm{m}^{-2}
\end{equation}
with about $95\%$ confidence.
This seemingly small value suggests the possible applicability of Theorem \ref{almostabelian} to the physical universe. Note, however, that since $\delta$ is not dimensionless any discussion of its `smallness' must come in comparison to something else. The theorem relates smallness to a diameter upper bound, volume lower bound, and modulus bound for $X$. Since the volume of $M^3$ and magnitude of $X$ are well-behaved, it is the diameter that should be used to interpret the smallness of $\delta$. Although an exact formula for the dependence of $\delta$ on the diameter is not known, a reasonable conjecture is that the dimensionless quantity $\delta \cdot\mathrm{diam}^2$ should be bounded above by a universal constant in order for the conclusions of the theorem to hold. Notice that the inverse squared diameter of the universe, as discussed in the previous section, is on par with the value in \eqref{tytyi}. In particular, if the diameter was known definitively to be slightly smaller than the proposed upper bounds, then the almost abelian fundamental group property would be confirmed.

\subsection{Consequences of Geometrization}

The geometrization theorem \cite{CaoZhu,KleinerLott,MorganTian,Perelman} gives a classification of compact 3-manifolds. Here we will work under the conclusion of Theorem \ref{almostabelian}, that the spatial slice $M^3$ has almost abelian fundamental group, and derive the possible topologies from the classification. The first observation \cite[Theorem 6]{KWW} is that this assumption implies that $M^3$ cannot be expressed as a connected sum $M_1\# M_2$, where $M_1$ and $M_2$ are compact manifolds having nontrivial fundamental groups, except possibly in the case that $\pi_1(M_1)=\pi_1(M_2)=\mathbb{Z}_2$. Thus either $M^3$ is prime, or $M^3=\mathbb{RP}^3 \# \mathbb{RP}^3$ is the connected sum of two copies of real projective space.

Let $N^3$ be a finite cover of $M^3$ with abelian fundamental group. Since $M^3$ is compact, $N^3$ is compact as well. Moreover, it may be assumed without loss of generality that $N^3$ is orientable, since we may take the orientable double cover if necessary. Geometrization may be used to provide a list \cite[Table 1.2]{AFW} of the possible abelian fundamental groups associated with orientable compact 3-manifolds. Namely, $\pi_1(N^3)$ may be one of the following groups: $\mathbb{Z}$, $\mathbb{Z}^3$, and $\mathbb{Z}_p$ the finite cyclic group of order $p$. Note that examples of 3-manifolds realizing these groups, respectively, are the ring $S^1 \times S^2$, the torus $T^3$, and the lens space $L(p,q)$ where $q$ is coprime to $p$.
In fact, we claim that $N^3$ must be one of these three types of manifolds. To see this, first observe that due to the restriction on the fundamental group $N^3$ must be prime. According to \cite[Theorem 2.1.2]{AFW}, if $N_1$ and $N_2$ are two closed orientable prime 3-manifolds with isomorphic fundamental groups, and they are not lens spaces, then they must be homeomorphic. It follows that if $\pi_1\left(N^3\right)=\mathbb{Z}$ then $N^3$ is homeomorphic to $S^1\times S^2$, whereas if $\pi_1\left(N^3\right)=\mathbb{Z}^3$ then $N^3$ is homeomorphic to $T^3$. The only other possibility is that $\pi_1\left(N^3\right)=\mathbb{Z}_p$. Since this is a finite group, we may use the elliptization portion of the geometrization theorem to conclude that $N^3$ is homeomorphic to a lens space $L(p,q)$.

The above arguments show that, when the conclusion of Theorem \ref{almostabelian} holds, $M^3$ must be covered by $S^3$, $S^1 \times S^2$, or $T^3$. In the second case, more can be said. In particular, if $M^3$ is covered by $S^1 \times S^2$ then it is also covered by the universal cover $\mathbb{R}\times S^2$. By \cite[Theorem 1]{Tol}, this implies that $M^3$ must be either: $S^1\times S^2$, the non-orientable 2-sphere bundle over the circle $S^1 \tilde{\times}S^2$, $S^1\times\mathbb{RP}^2$, or $\mathbb{RP}^3 \# \mathbb{RP}^3$.

\subsection{Conclusions}

We have analyzed the topology and other aspects of highly general cosmological models, which are completely divorced from the typical symmetry assumptions of homogeneity and isotropy. Conditions that guarantee a closed universe have been given, along with
a bound
for its diameter, using an extended version of the Bonnet-Myers theorem.
%
Further analysis of the topology under weaker hypotheses involving the interplay between curvature and diameter, derived from a generalized almost splitting theorem combined with geometrization of 3-manifolds, has provided a list of possible topologies for a closed universe. Specifically, the spatial sections allowed under Theorem \ref{almostabelian} must be either covered by $S^3$ or $T^3$, or are one of $S^1 \times S^2$, $S^1 \tilde{\times}S^2$, $S^1\times\mathbb{RP}^2$, or $\mathbb{RP}^3 \# \mathbb{RP}^3$. In particular, this eliminates connected sums (except for one) as well as the vast array of hyperbolic topologies.

Finally, we consider the estimate for $\delta$. The almost splitting theorem states that the critical value for $\delta$, below which topological conclusions can be drawn, is a function of parameters such as the diameter of the universe. However, the theorem does not construct this function, or provide an obvious way to directly analyze its properties. It is therefore an intriguing open problem to determine estimates for this critical value. Nevertheless, the fact that this value is nonzero implies that even with negative Ricci curvature of spatial sections, strong topological restrictions for the universe are still very much possible depending on the size of $\delta$. This is in sharp contrast to the standard model of cosmology, where only nonnegative scalar curvature of spatial slices induces such topological rigidity. Moreover, in further contrast to the standard model, these conclusions hold in highly general settings without the assumptions of homogeneity and isotropy, or in fact without any symmetry hypotheses at all.

\medskip

\noindent \textbf{Acknowledgements.}  The authors would like to thank George F. R. Ellis, Marilena Loverde, Dmitri Pogosyan, and Neelima Sehgal for helpful suggestions.

\end{document}